
%
%

%
%
%
%

\def\Serif{cmr}
\def\SerifBold{cmbx}
\def\SerifItalics{cmti}
\def\SerifSlanted{cmsl}
\def\SerifBoldItalics{cmbxti}
\def\SansSerif{cmss}
\def\SansSerifBold{cmssbx}
\def\SansSerifItalics{cmssi}
\def\SansSerifSlanted{cmssi}
\def\Math{cmmi}
\def\Symbols{cmsy}
\def\MathBold{cmmib}
\def\MoreSymbols{cmex}
\def\Typewriter{cmtt}
\def\Gothic{eufm}
\def\Double{msbm}
\def\Relazioni{msam}

= 			\Serif10 			at 5pt
= 		\SerifBold10 		at 5pt
= 	\SerifItalics10 	at 5pt
=		\SerifSlanted10 	at 5pt
=	\SerifBoldItalics10	at 5pt
= 		\SansSerif10 		at 5pt
=	\SansSerifBold10	at 5pt
=	\SansSerifItalics10	at 5pt
=	\SansSerifSlanted10	at 5pt
=				\Math10				at 5pt
=			\MathBold10			at 5pt
=			\Symbols10			at 5pt
=		\MoreSymbols10		at 5pt
=		\Typewriter10		at 5pt
=			\Gothic10			at 5pt
=			\Double10			at 5pt

= 			\Serif10 			at 7pt
= 		\SerifBold10 		at 7pt
= 	\SerifItalics10 	at 7pt
=	\SerifSlanted10 	at 7pt
=\SerifBoldItalics10	at 7pt
= 		\SansSerif10 		at 7pt
= 	\SansSerifBold10 	at 7pt
=\SansSerifItalics10	at 7pt
=\SansSerifSlanted10	at 7pt
=			\Math10				at 7pt
=		\MathBold10			at 7pt
=			\Symbols10			at 7pt
=		\MoreSymbols10		at 7pt
=		\Typewriter10		at 7pt
=			\Gothic10			at 7pt
=			\Double10			at 7pt

= 			\Serif10 			at 8pt
= 		\SerifBold10 		at 8pt
= 	\SerifItalics10 	at 8pt
=	\SerifSlanted10 	at 8pt
=\SerifBoldItalics10	at 8pt
= 		\SansSerif10 		at 8pt
= 	\SansSerifBold10 	at 8pt
=\SansSerifItalics10 at 8pt
=\SansSerifSlanted10 at 8pt
=			\Math10				at 8pt
=		\MathBold10			at 8pt
=			\Symbols10			at 8pt
=		\MoreSymbols10		at 8pt
=		\Typewriter10		at 8pt
=			\Gothic10			at 8pt
=			\Double10			at 8pt

= 			\Serif10 			at 10pt
= 		\SerifBold10 		at 10pt
= 		\SerifItalics10 	at 10pt
=		\SerifSlanted10 	at 10pt
=	\SerifBoldItalics10	at 10pt
= 		\SansSerif10 		at 10pt
= 	\SansSerifBold10 	at 10pt
= 	\SansSerifItalics10 at 10pt
= 	\SansSerifSlanted10 at 10pt
=				\Math10				at 10pt
=			\MathBold10			at 10pt
=			\Symbols10			at 10pt
=		\MoreSymbols10		at 10pt
=		\Typewriter10		at 10pt
=			\Gothic10			at 10pt
=			\Double10			at 10pt
=			\Relazioni10			at 10pt

= 				\Serif10 			at 12pt
= 			\SerifBold10 		at 12pt
= 		\SerifItalics10 	at 12pt
=		\SerifSlanted10 	at 12pt
=	\SerifBoldItalics10	at 12pt
= 			\SansSerif10 		at 12pt
= 		\SansSerifBold10 	at 12pt
= 	\SansSerifItalics10 at 12pt
= 	\SansSerifSlanted10 at 12pt
=				\Math10				at 12pt
=			\MathBold10			at 12pt
=			\Symbols10			at 12pt
=		\MoreSymbols10		at 12pt
=			\Typewriter10		at 12pt
=				\Gothic10			at 12pt
=				\Double10			at 12pt

= 			\Serif10 			at 14pt
= 		\SerifBold10 		at 14pt
= 	\SerifItalics10 	at 14pt
=		\SerifSlanted10 	at 14pt
=	\SerifBoldItalics10	at 14pt
= 		\SansSerif10 		at 14pt
= 	\SansSerifBold10 	at 14pt
= \SansSerifSlanted10 at 14pt
= \SansSerifItalics10 at 14pt
=				\Math10				at 14pt
=			\MathBold10			at 14pt
=			\Symbols10			at 14pt
=		\MoreSymbols10		at 14pt
=		\Typewriter10		at 14pt
=			\Gothic10			at 14pt
=			\Double10			at 14pt

\def\NormalStyle{\parindent=5pt\parskip=3pt\normalbaselineskip=14pt%
\def\nt{\tenSerif}%
\def\rm{\fam0\tenSerif}%
\textfont0=\tenSerif\scriptfont0=\sevenSerif\scriptscriptfont0=\fiveSerif
\textfont1=\tenMath\scriptfont1=\sevenMath\scriptscriptfont1=\fiveMath
\textfont2=\tenSymbols\scriptfont2=\sevenSymbols\scriptscriptfont2=\fiveSymbols
\textfont3=\tenMoreSymbols\scriptfont3=\sevenMoreSymbols\scriptscriptfont3=\fiveMoreSymbols
\textfont\itfam=\tenSerifItalics\def\it{\fam\itfam\tenSerifItalics}%
\textfont\slfam=\tenSerifSlanted\def\sl{\fam\slfam\tenSerifSlanted}%
\textfont\ttfam=\tenTypewriter\def\tt{\fam\ttfam\tenTypewriter}%
\textfont\bffam=\tenSerifBold%
\def\bf{\fam\bffam\tenSerifBold}\scriptfont\bffam=\sevenSerifBold\scriptscriptfont\bffam=\fiveSerifBold%
\def\cal{\tenSymbols}%
\def\greekbold{\tenMathBold}%
\def\gothic{\tenGothic}%
\def\Bbb{\tenDouble}%
\def\LieFont{\tenSerifItalics}%
\nt\normalbaselines\baselineskip=14pt%
}

\def\TitleStyle{\parindent=0pt\parskip=0pt\normalbaselineskip=15pt%
\def\nt{\fourteenSansSerifBold}%
\def\rm{\fam0\fourteenSansSerifBold}%
\textfont0=\fourteenSansSerifBold\scriptfont0=\tenSansSerifBold\scriptscriptfont0=\eightSansSerifBold
\textfont1=\fourteenMath\scriptfont1=\tenMath\scriptscriptfont1=\eightMath
\textfont2=\fourteenSymbols\scriptfont2=\tenSymbols\scriptscriptfont2=\eightSymbols
\textfont3=\fourteenMoreSymbols\scriptfont3=\tenMoreSymbols\scriptscriptfont3=\eightMoreSymbols
\textfont\itfam=\fourteenSansSerifItalics\def\it{\fam\itfam\fourteenSansSerifItalics}%
\textfont\slfam=\fourteenSansSerifSlanted\def\sl{\fam\slfam\fourteenSerifSansSlanted}%
\textfont\ttfam=\fourteenTypewriter\def\tt{\fam\ttfam\fourteenTypewriter}%
\textfont\bffam=\fourteenSansSerif%
\def\bf{\fam\bffam\fourteenSansSerif}\scriptfont\bffam=\tenSansSerif\scriptscriptfont\bffam=\eightSansSerif%
\def\cal{\fourteenSymbols}%
\def\greekbold{\fourteenMathBold}%
\def\gothic{\fourteenGothic}%
\def\Bbb{\fourteenDouble}%
\def\LieFont{\fourteenSerifItalics}%
\nt\normalbaselines\baselineskip=15pt%
}

\def\PartStyle{\parindent=0pt\parskip=0pt\normalbaselineskip=15pt%
\def\nt{\fourteenSansSerifBold}%
\def\rm{\fam0\fourteenSansSerifBold}%
\textfont0=\fourteenSansSerifBold\scriptfont0=\tenSansSerifBold\scriptscriptfont0=\eightSansSerifBold
\textfont1=\fourteenMath\scriptfont1=\tenMath\scriptscriptfont1=\eightMath
\textfont2=\fourteenSymbols\scriptfont2=\tenSymbols\scriptscriptfont2=\eightSymbols
\textfont3=\fourteenMoreSymbols\scriptfont3=\tenMoreSymbols\scriptscriptfont3=\eightMoreSymbols
\textfont\itfam=\fourteenSansSerifItalics\def\it{\fam\itfam\fourteenSansSerifItalics}%
\textfont\slfam=\fourteenSansSerifSlanted\def\sl{\fam\slfam\fourteenSerifSansSlanted}%
\textfont\ttfam=\fourteenTypewriter\def\tt{\fam\ttfam\fourteenTypewriter}%
\textfont\bffam=\fourteenSansSerif%
\def\bf{\fam\bffam\fourteenSansSerif}\scriptfont\bffam=\tenSansSerif\scriptscriptfont\bffam=\eightSansSerif%
\def\cal{\fourteenSymbols}%
\def\greekbold{\fourteenMathBold}%
\def\gothic{\fourteenGothic}%
\def\Bbb{\fourteenDouble}%
\def\LieFont{\fourteenSerifItalics}%
\nt\normalbaselines\baselineskip=15pt%
}

\def\ChapterStyle{\parindent=0pt\parskip=0pt\normalbaselineskip=15pt%
\def\nt{\fourteenSansSerifBold}%
\def\rm{\fam0\fourteenSansSerifBold}%
\textfont0=\fourteenSansSerifBold\scriptfont0=\tenSansSerifBold\scriptscriptfont0=\eightSansSerifBold
\textfont1=\fourteenMath\scriptfont1=\tenMath\scriptscriptfont1=\eightMath
\textfont2=\fourteenSymbols\scriptfont2=\tenSymbols\scriptscriptfont2=\eightSymbols
\textfont3=\fourteenMoreSymbols\scriptfont3=\tenMoreSymbols\scriptscriptfont3=\eightMoreSymbols
\textfont\itfam=\fourteenSansSerifItalics\def\it{\fam\itfam\fourteenSansSerifItalics}%
\textfont\slfam=\fourteenSansSerifSlanted\def\sl{\fam\slfam\fourteenSerifSansSlanted}%
\textfont\ttfam=\fourteenTypewriter\def\tt{\fam\ttfam\fourteenTypewriter}%
\textfont\bffam=\fourteenSansSerif%
\def\bf{\fam\bffam\fourteenSansSerif}\scriptfont\bffam=\tenSansSerif\scriptscriptfont\bffam=\eightSansSerif%
\def\cal{\fourteenSymbols}%
\def\greekbold{\fourteenMathBold}%
\def\gothic{\fourteenGothic}%
\def\Bbb{\fourteenDouble}%
\def\LieFont{\fourteenSerifItalics}%
\nt\normalbaselines\baselineskip=15pt%
}

\def\SectionStyle{\parindent=0pt\parskip=0pt\normalbaselineskip=13pt%
\def\nt{\twelveSansSerifBold}%
\def\rm{\fam0\twelveSansSerifBold}%
\textfont0=\twelveSansSerifBold\scriptfont0=\eightSansSerifBold\scriptscriptfont0=\eightSansSerifBold
\textfont1=\twelveMath\scriptfont1=\eightMath\scriptscriptfont1=\eightMath
\textfont2=\twelveSymbols\scriptfont2=\eightSymbols\scriptscriptfont2=\eightSymbols
\textfont3=\twelveMoreSymbols\scriptfont3=\eightMoreSymbols\scriptscriptfont3=\eightMoreSymbols
\textfont\itfam=\twelveSansSerifItalics\def\it{\fam\itfam\twelveSansSerifItalics}%
\textfont\slfam=\twelveSansSerifSlanted\def\sl{\fam\slfam\twelveSerifSansSlanted}%
\textfont\ttfam=\twelveTypewriter\def\tt{\fam\ttfam\twelveTypewriter}%
\textfont\bffam=\twelveSansSerif%
\def\bf{\fam\bffam\twelveSansSerif}\scriptfont\bffam=\eightSansSerif\scriptscriptfont\bffam=\eightSansSerif%
\def\cal{\twelveSymbols}%
\def\bg{\twelveMathBold}%
\def\gothic{\twelveGothic}%
\def\Bbb{\twelveDouble}%
\def\LieFont{\twelveSerifItalics}%
\nt\normalbaselines\baselineskip=13pt%
}

\def\SubSectionStyle{\parindent=0pt\parskip=0pt\normalbaselineskip=13pt%
\def\nt{\twelveSansSerifItalics}%
\def\rm{\fam0\twelveSansSerifItalics}%
\textfont0=\twelveSansSerifItalics\scriptfont0=\eightSansSerifItalics\scriptscriptfont0=\eightSansSerifItalics%
\textfont1=\twelveMath\scriptfont1=\eightMath\scriptscriptfont1=\eightMath%
\textfont2=\twelveSymbols\scriptfont2=\eightSymbols\scriptscriptfont2=\eightSymbols%
\textfont3=\twelveMoreSymbols\scriptfont3=\eightMoreSymbols\scriptscriptfont3=\eightMoreSymbols%
\textfont\itfam=\twelveSansSerif\def\it{\fam\itfam\twelveSansSerif}%
\textfont\slfam=\twelveSansSerifSlanted\def\sl{\fam\slfam\twelveSerifSansSlanted}%
\textfont\ttfam=\twelveTypewriter\def\tt{\fam\ttfam\twelveTypewriter}%
\textfont\bffam=\twelveSansSerifBold%
\def\bf{\fam\bffam\twelveSansSerifBold}\scriptfont\bffam=\eightSansSerifBold\scriptscriptfont\bffam=\eightSansSerifBold%
\def\cal{\twelveSymbols}%
\def\greekbold{\twelveMathBold}%
\def\gothic{\twelveGothic}%
\def\Bbb{\twelveDouble}%
\def\LieFont{\twelveSerifItalics}%
\nt\normalbaselines\baselineskip=13pt%
}

\def\AuthorStyle{\parindent=0pt\parskip=0pt\normalbaselineskip=14pt%
\def\nt{\tenSerif}%
\def\rm{\fam0\tenSerif}%
\textfont0=\tenSerif\scriptfont0=\sevenSerif\scriptscriptfont0=\fiveSerif
\textfont1=\tenMath\scriptfont1=\sevenMath\scriptscriptfont1=\fiveMath
\textfont2=\tenSymbols\scriptfont2=\sevenSymbols\scriptscriptfont2=\fiveSymbols
\textfont3=\tenMoreSymbols\scriptfont3=\sevenMoreSymbols\scriptscriptfont3=\fiveMoreSymbols
\textfont\itfam=\tenSerifItalics\def\it{\fam\itfam\tenSerifItalics}%
\textfont\slfam=\tenSerifSlanted\def\sl{\fam\slfam\tenSerifSlanted}%
\textfont\ttfam=\tenTypewriter\def\tt{\fam\ttfam\tenTypewriter}%
\textfont\bffam=\tenSerifBold%
\def\bf{\fam\bffam\tenSerifBold}\scriptfont\bffam=\sevenSerifBold\scriptscriptfont\bffam=\fiveSerifBold%
\def\cal{\tenSymbols}%
\def\greekbold{\tenMathBold}%
\def\gothic{\tenGothic}%
\def\Bbb{\tenDouble}%
\def\LieFont{\tenSerifItalics}%
\nt\normalbaselines\baselineskip=14pt%
}

\def\AddressStyle{\parindent=0pt\parskip=0pt\normalbaselineskip=14pt%
\def\nt{\eightSerif}%
\def\rm{\fam0\eightSerif}%
\textfont0=\eightSerif\scriptfont0=\sevenSerif\scriptscriptfont0=\fiveSerif
\textfont1=\eightMath\scriptfont1=\sevenMath\scriptscriptfont1=\fiveMath
\textfont2=\eightSymbols\scriptfont2=\sevenSymbols\scriptscriptfont2=\fiveSymbols
\textfont3=\eightMoreSymbols\scriptfont3=\sevenMoreSymbols\scriptscriptfont3=\fiveMoreSymbols
\textfont\itfam=\eightSerifItalics\def\it{\fam\itfam\eightSerifItalics}%
\textfont\slfam=\eightSerifSlanted\def\sl{\fam\slfam\eightSerifSlanted}%
\textfont\ttfam=\eightTypewriter\def\tt{\fam\ttfam\eightTypewriter}%
\textfont\bffam=\eightSerifBold%
\def\bf{\fam\bffam\eightSerifBold}\scriptfont\bffam=\sevenSerifBold\scriptscriptfont\bffam=\fiveSerifBold%
\def\cal{\eightSymbols}%
\def\greekbold{\eightMathBold}%
\def\gothic{\eightGothic}%
\def\Bbb{\eightDouble}%
\def\LieFont{\eightSerifItalics}%
\nt\normalbaselines\baselineskip=14pt%
}

\def\AbstractStyle{\parindent=0pt\parskip=0pt\normalbaselineskip=12pt%
\def\nt{\eightSerif}%
\def\rm{\fam0\eightSerif}%
\textfont0=\eightSerif\scriptfont0=\sevenSerif\scriptscriptfont0=\fiveSerif
\textfont1=\eightMath\scriptfont1=\sevenMath\scriptscriptfont1=\fiveMath
\textfont2=\eightSymbols\scriptfont2=\sevenSymbols\scriptscriptfont2=\fiveSymbols
\textfont3=\eightMoreSymbols\scriptfont3=\sevenMoreSymbols\scriptscriptfont3=\fiveMoreSymbols
\textfont\itfam=\eightSerifItalics\def\it{\fam\itfam\eightSerifItalics}%
\textfont\slfam=\eightSerifSlanted\def\sl{\fam\slfam\eightSerifSlanted}%
\textfont\ttfam=\eightTypewriter\def\tt{\fam\ttfam\eightTypewriter}%
\textfont\bffam=\eightSerifBold%
\def\bf{\fam\bffam\eightSerifBold}\scriptfont\bffam=\sevenSerifBold\scriptscriptfont\bffam=\fiveSerifBold%
\def\cal{\eightSymbols}%
\def\greekbold{\eightMathBold}%
\def\gothic{\eightGothic}%
\def\Bbb{\eightDouble}%
\def\LieFont{\eightSerifItalics}%
\nt\normalbaselines\baselineskip=12pt%
}

\def\RefsStyle{\parindent=0pt\parskip=0pt%
\def\nt{\eightSerif}%
\def\rm{\fam0\eightSerif}%
\textfont0=\eightSerif\scriptfont0=\sevenSerif\scriptscriptfont0=\fiveSerif
\textfont1=\eightMath\scriptfont1=\sevenMath\scriptscriptfont1=\fiveMath
\textfont2=\eightSymbols\scriptfont2=\sevenSymbols\scriptscriptfont2=\fiveSymbols
\textfont3=\eightMoreSymbols\scriptfont3=\sevenMoreSymbols\scriptscriptfont3=\fiveMoreSymbols
\textfont\itfam=\eightSerifItalics\def\it{\fam\itfam\eightSerifItalics}%
\textfont\slfam=\eightSerifSlanted\def\sl{\fam\slfam\eightSerifSlanted}%
\textfont\ttfam=\eightTypewriter\def\tt{\fam\ttfam\eightTypewriter}%
\textfont\bffam=\eightSerifBold%
\def\bf{\fam\bffam\eightSerifBold}\scriptfont\bffam=\sevenSerifBold\scriptscriptfont\bffam=\fiveSerifBold%
\def\cal{\eightSymbols}%
\def\greekbold{\eightMathBold}%
\def\gothic{\eightGothic}%
\def\Bbb{\eightDouble}%
\def\LieFont{\eightSerifItalics}%
\nt\normalbaselines\baselineskip=10pt%
}



%
%


\def\ModeYes{yes}
\def\ModeNo{no}

\def\ModeUndef{undefined}


\def\nx{\noexpand}
\def\ni{\noindent}
\def\newpage{\vfill\eject}

\def\ss{\vskip 5pt}
\def\ms{\vskip 10pt}
\def\bs{\vskip 20pt}

 \def\,{\mskip\thinmuskip}
 \def\!{\mskip-\thinmuskip}
 \def\>{\mskip\medmuskip}
 \def\;{\mskip\thickmuskip}

%
% ... 
%

\def\refsModePost{post}
\def\refsModeAuto{auto}

\def\dbRefsSatusModeOk{ok}
\def\dbRefsSatusModeError{error}
\def\dbRefsSatusModeWarning{warning}


\newcount\BNUM
\BNUM=0

\def\refs{}

\def\SetModePost{\xdef\refsMode{\refsModePost}}			
\SetModePost

\def\dbRefsStatusOk{%
	\xdef\dbRefsStatus{\dbRefsSatusModeOk}%
	\xdef\dbRefsError{\ModeNo}%
	\xdef\dbRefsWarning{\ModeNo}%
	\xdef\dbRefsInfo{\ModeNo}%
}

\def\dbRefs{%
}

\def\dbRefsGet#1{%
	\xdef\found{N}\xdef\ikey{#1}\dbRefsStatusOk%
	\xdef\key{\ModeUndef}\xdef\tag{\ModeUndef}\xdef\tail{\ModeUndef}%
	\dbRefs%
}

\def\NextRefsTag{%
	\global\advance\BNUM by 1%
}
\def\ShowTag#1{{\bf [#1]}}

\def\dbRefsInsert#1#2{%
\dbRefsGet{#1}%
\if\found Y %
   \xdef\dbRefsStatus{\dbRefsSatusModeWarning}%
   \xdef\dbRefsWarning{record is already there}%
   \xdef\dbRefsInfo{record not inserted}%
\else%
   \toks2=\expandafter{\dbRefs}%
   \ifx\refsMode\refsModeAuto \NextRefsTag
    \xdef\dbRefs{%
   	\the\toks2 \nx\xdef\nx\dbx{#1}%
	\nx\ifx\nx\ikey %
		\nx\dbx\nx\xdef\nx\found{Y}%
		\nx\xdef\nx\key{#1}%
		\nx\xdef\nx\tag{\the\BNUM}%
		\nx\xdef\nx\tail{#2}%
	\nx\fi}%
	\global\xdef\refs{\refs \ss\ni[\the\BNUM]\ #2\par}
   \fi%
   \ifx\refsMode\refsModePost 
    \xdef\dbRefs{%
   	\the\toks2 \nx\xdef\nx\dbx{#1}%
	\nx\ifx\nx\ikey %
		\nx\dbx\nx\xdef\nx\found{Y}%
		\nx\xdef\nx\key{#1}%
		\nx\xdef\nx\tag{\ModeUndef}%
		\nx\xdef\nx\tail{#2}%
	\nx\fi}%
   \fi%
\fi%
}

\def\dbRefsEdit#1#2#3{\dbRefsGet{#1}%
\if\found N 
   \xdef\dbRefsStatus{\dbRefsSatusModeError}%
   \xdef\dbRefsError{record is not there}%
   \xdef\dbRefsInfo{record not edited}%
\else%
   \toks2=\expandafter{\dbRefs}%
   \xdef\dbRefs{\the\toks2%
   \nx\xdef\nx\dbx{#1}%
   \nx\ifx\nx\ikey\nx\dbx %
	\nx\xdef\nx\found{Y}%
	\nx\xdef\nx\key{#1}%
	\nx\xdef\nx\tag{#2}%
	\nx\xdef\nx\tail{#3}%
   \nx\fi}%
\fi%
}

\def\bib#1#2{\RefsStyle\dbRefsInsert{#1}{#2}%
	\ifx\dbRefsStatus\dbRefsSatusModeWarning %
		\message{^^J}%
		\message{WARNING: Reference [#1] is doubled.^^J}%
	\fi%
}

\def\ref#1{\dbRefsGet{#1}%
\ifx\found N %
  \message{^^J}%
  \message{ERROR: Reference [#1] unknown.^^J}%
  \ShowTag{??}%
\else%
	\ifx\tag\ModeUndef \NextRefsTag%
		\dbRefsEdit{#1}{\the\BNUM}{\tail}%
		\dbRefsGet{#1}%
		\global\xdef\refs{\refs \ss\ni [\tag]\ \tail\par}
	\fi
	\ShowTag{\tag}%
\fi%
}

\def\ShowBiblio{\ms\Ensure{\SectionEnsure}%
{\SectionStyle\ni References}%
{\RefsStyle\refs}%
}

\newcount\CHANGES
\CHANGES=0
\def\AuxFile{7}
\def\PreventDoubleOn{\xdef\PreventDoubleLabel{\ModeYes}}

\PreventDoubleOn

\def\StoreLabel#1#2{\xdef\itag{#2}
 \ifx\PreModeStatus\ModeNo %
   \message{^^J}%
   \errmessage{You can't use Check without starting with OpenPreMode (and finishing with
ClosePreMode)^^J}%
 \else%
   \immediate\write\AuxFile{\nx\dbLabelPreInsert{#1}{\itag}}%
   \dbLabelGet{#1}%
   \ifx\itag\tag %
   \else%
	\global\advance\CHANGES by 1%
 	\xdef\itag{(?.??)}%
    \fi%
   \fi%
}

\def\PreModeStatus{\ModeNo}

\def\ClosePreMode{\immediate\closeout\AuxFile%
  \ifnum\CHANGES=0%
	\message{^^J}%
	\message{**********************************^^J}%
	\message{**  NO CHANGES TO THE AuxFile  **^^J}%
	\message{**********************************^^J}%
 \else%
	\message{^^J}%
	\message{**************************************************^^J}%
	\message{**  PLAEASE TYPESET IT AGAIN (\the\CHANGES)  **^^J}%
    \errmessage{**************************************************^^ J}%
  \fi%
  \edef\PreModeStatus{\ModeNo}%
}

\def\dbLabelSatusModeOk{ok}

\def\dbLabelSatusModeWarning{warning}

\def\dbLabelStatusOk{%
	\xdef\dbLabelStatus{\dbLabelSatusModeOk}%
	\xdef\dbLabelError{\ModeNo}%
	\xdef\dbLabelWarning{\ModeNo}%
	\xdef\dbLabelInfo{\ModeNo}%
}

\def\dbLabel{%
}

\def\dbLabelGet#1{%
	\xdef\found{N}\xdef\ikey{#1}\dbLabelStatusOk%
	\xdef\key{\ModeUndef}\xdef\tag{\ModeUndef}\xdef\pre{\ModeUndef}%
	\dbLabel%
}

\def\ShowLabel#1{%
 \dbLabelGet{#1}%
 \ifx\tag \ModeUndef %
 	\global\advance\CHANGES by 1%
 	(?.??)%
 \else%
 	\tag%
 \fi%
}

\def\dbLabelPreInsert#1#2{\dbLabelGet{#1}%
\if\found Y %
  \xdef\dbLabelStatus{\dbLabelSatusModeWarning}%
   \xdef\dbLabelWarning{Label is already there}%
   \xdef\dbLabelInfo{Label not inserted}%
   \message{^^J}%
   \errmessage{Double pre definition of label [#1]^^J}%
\else%
   \toks2=\expandafter{\dbLabel}%
    \xdef\dbLabel{%
   	\the\toks2 \nx\xdef\nx\dbx{#1}%
	\nx\ifx\nx\ikey %
		\nx\dbx\nx\xdef\nx\found{Y}%
		\nx\xdef\nx\key{#1}%
		\nx\xdef\nx\tag{#2}%
		\nx\xdef\nx\pre{\ModeYes}%
	\nx\fi}%
\fi%
}

\def\dbLabelInsert#1#2{\dbLabelGet{#1}%
\xdef\itag{#2}%
\dbLabelGet{#1}%
\if\found Y %
	\ifx\tag\itag %
	\else%
	   \ifx\PreventDoubleLabel\ModeYes %
		\message{^^J}%
		\errmessage{Double definition of label [#1]^^J}%
	   \else%
		\message{^^J}%
		\message{Double definition of label [#1]^^J}%
	   \fi%
	\fi%
   \xdef\dbLabelStatus{\dbLabelSatusModeWarning}%
   \xdef\dbLabelWarning{Label is already there}%
   \xdef\dbLabelInfo{Label not inserted}%
\else%
   \toks2=\expandafter{\dbLabel}%
    \xdef\dbLabel{%
   	\the\toks2 \nx\xdef\nx\dbx{#1}%
	\nx\ifx\nx\ikey %
		\nx\dbx\nx\xdef\nx\found{Y}%
		\nx\xdef\nx\key{#1}%
		\nx\xdef\nx\tag{#2}%
		\nx\xdef\nx\pre{\ModeNo}%
	\nx\fi}%
\fi%
}


\newcount\PART
\newcount\CHAPTER
\newcount\SECTION
\newcount\SUBSECTION
\newcount\FNUMBER

\PART=0
\CHAPTER=0
\SECTION=0
\SUBSECTION=0	
\FNUMBER=0

\def\LastPart{\ModeUndef}
\def\LastChapter{\ModeUndef}
\def\LastSection{\ModeUndef}
\def\LastSubSection{\ModeUndef}
\def\LastClaim{\ModeUndef}
\def\Last{\ModeUndef}

\newdimen\TOBOTTOM
\newdimen\LIMIT

\def\Ensure#1{\ \par\ \immediate\LIMIT=#1\immediate\TOBOTTOM=\the\pagegoal\advance\TOBOTTOM by
-\pagetotal%
\ifdim\TOBOTTOM<\LIMIT\newpage \else%
\vskip-\parskip\vskip-\parskip\vskip-\baselineskip\fi}

\def\PartLabel{\the\PART}
\def\NewPart#1{\global\advance\PART by 1%
         \bs\ni{\PartStyle  Part \PartLabel:}
         \bs\ni{\PartStyle #1}\newpage%
         \CHAPTER=0\SECTION=0\SUBSECTION=0\FNUMBER=0%
         \gdef\Left{#1}%
         \global\edef\Last{\PartLabel}%
         \global\edef\LastPart{\PartLabel}%
         \global\edef\LastChapter{\ModeUndef}%
         \global\edef\LastSection{\ModeUndef}%
         \global\edef\LastSubSection{\ModeUndef}%
         \global\edef\LastClaim{\ModeUndef}}
\def\ChapterLabel{\the\CHAPTER}
\def\NewChapter#1{\global\advance\CHAPTER by 1%
         \bs\ni{\ChapterStyle  Chapter \ChapterLabel: #1}\ms%
         \SECTION=0\SUBSECTION=0\FNUMBER=0%
         \gdef\Left{#1}%
         \global\edef\Last{\ChapterLabel}%
         \global\edef\LastChapter{\ChapterLabel}%
         \global\edef\LastSection{\ModeUndef}%
         \global\edef\LastSubSection{\ModeUndef}%
         \global\edef\LastClaim{\ModeUndef}}
\def\SectionEnsure{3cm}
\def\NewSection#1{\Ensure{\SectionEnsure}\gdef\SectionLabel{\the\SECTION}\global\advance\SECTION by
1%
         \ms\ni{\SectionStyle  \SectionLabel.\ #1}\ss%
         \SUBSECTION=0\FNUMBER=0%
         \gdef\Left{#1}%
         \global\edef\Last{\SectionLabel}%
         \global\edef\LastSection{\SectionLabel}%
         \global\edef\LastSubSection{\ModeUndef}%
         \global\edef\LastClaim{\ModeUndef}}
\def\NewAppendix#1#2{\Ensure{\SectionEnsure}\gdef\SectionLabel{#1}\global\advance\SECTION by 1%
         \bs\ni{\SectionStyle  Appendix \SectionLabel.\ #2}\ss%
         \SUBSECTION=0\FNUMBER=0%
         \gdef\Left{#2}%
         \global\edef\Last{\SectionLabel}%
         \global\edef\LastSection{\SectionLabel}%
         \global\edef\LastSubSection{\ModeUndef}%
         \global\edef\LastClaim{\ModeUndef}}
\def\Acknowledgements{\Ensure{\SectionEnsure}\gdef\SectionLabel{}%
         \ms\ni{\SectionStyle  Acknowledgments}\ss%
         \SECTION=0\SUBSECTION=0\FNUMBER=0%
         \gdef\Left{}%
         \global\edef\Last{\ModeUndef}%
         \global\edef\LastSection{\ModeUndef}%
         \global\edef\LastSubSection{\ModeUndef}%
         \global\edef\LastClaim{\ModeUndef}}
\def\SubSectionEnsure{2cm}
\def\SubSectionLabel{\ifnum\SECTION>0 \the\SECTION.\fi\the\SUBSECTION}
\def\NewSubSection#1{\Ensure{\SubSectionEnsure}\global\advance\SUBSECTION by 1%
         \ms\ni{\SubSectionStyle #1}\ss%
         \global\edef\Last{\SubSectionLabel}%
         \global\edef\LastSubSection{\SubSectionLabel}}
\def\SetNumberingModeN{\def\ClaimLabel{(\the\FNUMBER)}}
\def\SetNumberingModeSN{\def\ClaimLabel{(\ifnum\SECTION>0 \SectionLabel.\fi%
      \the\FNUMBER)}}
\def\SetNumberingModeCSN{\def\ClaimLabel{(\ifnum\CHAPTER>0 \the\CHAPTER.\fi%
      \ifnum\SECTION>0 \SectionLabel.\fi%
      \the\FNUMBER)}}

\def\NewClaim{\global\advance\FNUMBER by 1%
    \ClaimLabel%
    \global\edef\LastClaim{\ClaimLabel}%
    \global\edef\Last{\ClaimLabel}}

\def\HideLabels{\xdef\ShowLabelsMode{\ModeNo}}
\HideLabels

\def\fn{\eqno{\NewClaim}} 
\def\fl#1{%
\ifx\ShowLabelsMode\ModeYes%
 \eqno{{\buildrel{\hbox{\AbstractStyle[#1]}}\over{\hfill\NewClaim}}}%
\else%
 \eqno{\NewClaim}%
\fi%
\dbLabelInsert{#1}{\ClaimLabel}}
\def\fprel#1{\global\advance\FNUMBER by 1\StoreLabel{#1}{\ClaimLabel}%
\ifx\ShowLabelsMode\ModeYes%
\eqno{{\buildrel{\hbox{\AbstractStyle[#1]}}\over{\hfill.\itag}}}%
\else%
 \eqno{\itag}%
\fi%
}

\def\cl#1{\global\advance\FNUMBER by 1\dbLabelInsert{#1}{\ClaimLabel}%
\ifx\ShowLabelsMode\ModeYes%
${\buildrel{\hbox{\AbstractStyle[#1]}}\over{\hfill\ClaimLabel}}$%
\else%
  $\ClaimLabel$%
\fi%
}
\def\cprel#1{\global\advance\FNUMBER by 1\StoreLabel{#1}{\ClaimLabel}%
\ifx\ShowLabelsMode\ModeYes%
${\buildrel{\hbox{\AbstractStyle[#1]}}\over{\hfill.\itag}}$%
\else%
  $\itag$%
\fi%
}

\def\Note{\ms\leftskip 3cm\rightskip 1.5cm\AbstractStyle}
\def\endNote{\par\leftskip 2cm\rightskip 0cm\NormalStyle\ss}


\parindent=7pt
\leftskip=2cm
\newcount\SideIndent
\newcount\SideIndentTemp
\SideIndent=0
\newdimen\SectionIndent
\SectionIndent=-8pt

\def\sidebar{\vrule height15pt width.2pt }
\def\endcorner{\hbox{\hbox{\vrule height6pt width.2pt}\vbox to6pt{\vfill\hbox
to4pt{\leaders\hrule height0.2pt\hfill}}}}
\def\begincorner{\hbox{\hbox{\vrule height6pt width.2pt}\vbox to6pt{\hbox
to4pt{\leaders\hrule height0.2pt\hfill}}}}
\def\endbegincorner{\hbox{\vbox to15pt{\endcorner\vskip-6pt\begincorner\vfill}}}
\def\SideShow{\SideIndentTemp=\SideIndent \ifnum \SideIndentTemp>0 
\loop\sidebar\hskip 2pt \advance\SideIndentTemp by-1\ifnum \SideIndentTemp>1 \repeat\fi}

\def\BeginSection{{\vbadness 100000 \par\ni\hskip\SectionIndent%
\SideShow\vbox to 15pt{\vfill\begincorner}}\global\advance\SideIndent by1\vskip-10pt}

\def\EndSection{{\vbadness 100000 \par\ni\global\advance\SideIndent by-1%
\hskip\SectionIndent\SideShow\vbox to15pt{\endcorner\vfill}\vskip-10pt}}

\def\EndBeginSection{{\vbadness 100000\par\ni%
\global\advance\SideIndent by-1\hskip\SectionIndent\SideShow
\vbox to15pt{\vfill\endbegincorner}}%
\global\advance\SideIndent by1\vskip-10pt}

\def\ShowBeginCorners#1{%
\SideIndentTemp =#1 \advance\SideIndentTemp by-1%
\ifnum \SideIndentTemp>0 %
\vskip-15truept\hbox{\kern 2truept\vbox{\hbox{\begincorner}%
\ShowBeginCorners{\SideIndentTemp}\vskip-3truept}}%
\fi%
}

\def\ShowEndCorners#1{%
\SideIndentTemp =#1 \advance\SideIndentTemp by-1%
\ifnum \SideIndentTemp>0 %
\vskip-15truept\hbox{\kern 2truept\vbox{\hbox{\endcorner}%
\ShowEndCorners{\SideIndentTemp}\vskip 2truept}}%
\fi%
}

\def\BeginSections#1{{\vbadness 100000 \par\ni\hskip\SectionIndent%
\SideShow\vbox to 15pt{\vfill\ShowBeginCorners{#1}}}\global\advance\SideIndent by#1\vskip-10pt}

\def\EndSections#1{{\vbadness 100000 \par\ni\global\advance\SideIndent by-#1%
\hskip\SectionIndent\SideShow\vbox to15pt{\vskip15pt\ShowEndCorners{#1}\vfill}\vskip-10pt}}

\def\EndBeginSections#1#2{{\vbadness 100000\par\ni%
\global\advance\SideIndent by-#1%
\hbox{\hskip\SectionIndent\SideShow\kern-2pt%
\vbox to15pt{\vskip15pt\ShowEndCorners{#1}\vskip4pt\ShowBeginCorners{#2}}}}%
\global\advance\SideIndent by#2\vskip-10pt}




%
%


\def\al{\alpha}
\def\be{\beta}
\def\de{\delta}

\def\la{\lambda}

\def\om{\omega}
\def\si{\sigma}

\def\Ga{\Gamma}

\def\La{\Lambda}
\def\Om{\Omega}






\def\ip{\hbox to4pt{\leaders\hrule height0.3pt\hfill}\vbox to8pt{\leaders\vrule
width0.3pt\vfill}\kern 2pt}

\def\na{\nabla}

\def\then{\Rightarrow}

%
%

\def\cases#1{\left\{\eqalign{#1}\right.}
\NormalStyle
\SetNumberingModeSN
\PreventDoubleOn

\long\def\title#1{\centerline{\TitleStyle\ni#1}}

\long\def\author#1{\ms\centerline{\AuthorStyle by {\it #1}}}

\long\def\address#1{\ss\centerline{\AddressStyle #1}\par}
\long\def\moreaddress#1{\centerline{\AddressStyle #1}\par}
\def\abstract{\ms\leftskip 3cm\rightskip .5cm\AbstractStyle{\bf \ni Abstract:}\ }
\def\endabstract{\par\leftskip 2cm\rightskip 0cm\NormalStyle\ss}

\SetNumberingModeSN

\def\frac[#1/#2]{\hbox{$#1\over#2$}}
\def\Frac[#1/#2]{{#1\over#2}}
\def\({\left(}
\def\){\right)}
\def\[{\left[}
\def\]{\right]}
\def\^#1{{}^{#1}_{\>\cdot}}
\def\_#1{{}_{#1}^{\>\cdot}}
\def\Label=#1{{\buildrel {\hbox{\fiveSerif \ShowLabel{#1}}}\over =}}
\def\<{\kern -1pt}


\def\ExpandAllCNotes{\long\def\CNote##1{%
\BeginSection
	\Note%
 		##1%
	\endNote%
\EndSection%
}}
\ExpandAllCNotes
%
%
%
%


\def\frame#1{\vbox{\hrule\hbox{\vrule\vbox{\kern2pt\hbox{\kern2pt#1\kern2pt}\kern2pt}\vrule}\hrule\kern-4pt}}

\def\Box to #1#2#3{\frame{\vtop{\hbox to #1{\hfill #2 \hfill}\hbox to #1{\hfill #3 \hfill}}}}


\bib{Faraoni}{T.P. Sotiriou, V.Faraoni, 
{\it $f(R)$ theories of gravity},
arXiv:0805.1726}

\bib{HawkingEllis}{S.W.Hawking, G.F.R.Ellis, P.V.Landshoff, D.R.Nelson, D.W.Sciama, S.Weinberg,
{\it The Large Scale Structure of Space-Time},
Cambridge University Press, New York, 1973.
}

\bib{Wald}{R.M.Wald,
{\it General Relativity},
University of Chicago Press, 1984.
}

\bib{Capozzo}{S. Capozziello, M. De Laurentis, V. Faraoni
{\it A bird's eye view of $f(R)$-gravity}
(2009); arXiv:0909.4672}

\bib{Gibbons}{S.W.Hawking, C.J.Hunter, 
{\it Gravitational Entropy and Global Structure},
Phys.Rev. D59 (1999) 044025;
arXiv:hep-th/9808085v2
}

\bib{FFCov}{M.Ferraris, M.Francaviglia,
{\it Covariant first-order Lagrangians, energy-density and superpotentials in general relativity},
GRG {\bf 22}(9) (1990) 965-985}

\bib{Augmented}{L.Fatibene, M. Ferraris, M. Francaviglia,
{\it Augmented Variational Principles and Relative Conservation
Laws in Classical Field Theory},
Int. J. Geom. Methods Mod. Phys. {\bf 2}(3) (2005),  373-392}

\bib{Caroll}{S.M.Carrol, 
{\it Lecture Notes on General Relativity},
arXiv:gr-qc/9712019v1
}

\bib{tHooft}{G.'t Hooft,
{\it Introduction to General Relativity},
Rinton Press,. 2001.
}



\def\ubal{\underline{\al}\kern1pt}
\def\obal{\overline{\al}\kern1pt}

\def\ubR{\underline{R}\kern1pt}
\def\obR{\overline{R}\kern1pt}
\def\ubom{\underline{\om}\kern1pt}
\def\obxi{\overline{\xi}\kern1pt}
\def\ubu{\underline{u}\kern1pt}
\def\ube{\underline{e}\kern1pt}
\def\obe{\overline{e}\kern1pt}

\NormalStyle

\title{About Boundary Terms in Higher Order Theories}

\author{L.~Fatibene$^{a, b}$, M.~Francaviglia$^{a,b,c}$,
S.~Mercadante$^{a,b}$\footnote{}{\AbstractStyle eMail: {\tt lorenzo.fatibene@unito.it}, 
{\tt mauro.francaviglia@unito.it}, {\tt silvio.mercadante@unito.it}}}

\address{$^a$ Department of Mathematics, University of Torino (Italy)}

\moreaddress{$^b$ INFN - Iniziativa Specifica Na12}

\moreaddress{$^c$ LCS, University of Calabria (Italy)}

\abstract
It is shown that when in a higher order variational principle one fixes fields at the boundary leaving the field
derivatives unconstrained, then the variational principle (in particular the solution space) is not
invariant with respect to the addition of boundary terms to the action, as it happens instead when
the correct procedure is applied.  Examples are considered to show how leaving derivatives of fields
unconstrained affects the physical interpretation of the model.  This is justified in particularl by
the need of clarifying the issue for the purpose of applications to relativistic gravitational
theories, where a bit of confusion still exists.
On the contrary, as it is well known for variational principles of order $k$, if one fixes variables
together with their derivatives (up to order $k-1$) on the boundary then boundary terms leave
solution space invariant.  \endabstract

\NewSection{Introduction}

Recently the interest in higher order Lagrangian theories has been renewed within the framework of
covariant field theories in various contexts, aiming to suitably extend standard (Hilbert-Einstein)
General Relativity in order to model, at least partially, dark energy/matter effects (see
\ref{Capozzo} and references quoted therein) via the use of gravitational Lagrangians depending 
non-linearly on the curvature.

In gravitational literature different attitudes towards boundary conditions in GR and in alternative
gravitational theories are presented (see \ref{Faraoni} for a detailed review).  We shall here
stress that mathematical consequences of different attitudes must be considered {\sl before\/} any
physical interpretation is attempted and that of course one is not free to ignore these
consequences, that might be (and usually are) rather crucial for a number of physically relevant
issues, e.g.\ the definition of conservation laws and their correct physical interpretation.

From the mathematical viewpoint, any attitude towards boundary conditions should be dictated by
Hamilton's least action principle.  This principle is a {\sl definition\/} of the critical sections
which have to be understood as physical configurations.  Being it a {\sl definition\/} one is
logically free to choose the formulation which is more suitable to the situation.  However, there
are physical and mathematical consequences of this choice which must be in any case taken into
account.  Moreover, it would be appreciated if a general guiding principle would avoid to treat each
model on its own on the base of physical considerations which in some cases (e.g.~when dealing with
exotic physics or non-trivial generalizations of the models already considered) could be unclear.

In particular, we shall hereafter show that if one assumes that only the value of fields must be
fixed while (higher order) derivatives are left unconstrained at the boundary, then one cannot keep
that pure divergencies in the action leave the solution space invariant, as it happens in the
standard applications of Calculus of Variations.  This is particularly relevant for Gravity, since
in the literature (see e.g.\ \ref{Wald}) it is often claimed that in standard GR one is free to
choose not to fix first derivatives of the metric at the boundary, since the boundary terms of the
Hilbert action can be written as a total variation and hence can be compensated in various
non-unique ways by adding suitable boundary terms to the action.  Even if this is mathematically
correct in GR it is in any case rather misleading since such a procedure fails to hold if one
considers Lagrangians that are non-degenerate and non-linear in curvature.  Accordingly we believe
that whenever such a choice is adopted one should clearly state that this is done at the expense of
{\sl changing\/} the space of solutions and affecting conservation laws which is unfortunately 
physically disturbing; see also \ref{Gibbons}.
 
As a motivation for such an uncanonical choice it is often claimed that fixing higher order
variations of the fields may affect their physical interpretation so that this standard attitude
should not be embraced without considering these effects.  This is of course true and we fully agree
that detailed discussions on the role that different boundary conditions have in GR is extremely
important.  However, it is also true the other way around, i.e.\ when leaving variations of field 
derivatives free at the boundary one should always be careful about the change of solution space, 
the interpretation of boundary fluxes as well as the further spurious boundary equations that 
appear besides the (bulk) field equations.

Hereafter, we shall present explicit examples in Mechanics and Field Theory.  From these examples it
is clearly shown that if one artificially wants to describe a system by a higher order Lagrangian
adding pure divergencies to the Lagrangian itself, then in order to maintain the standard
interpretation of the physical system one is forced to fix variations {\sl and their derivatives\/} at
the boundary. The examples will in fact show, {\it en passant}, how the solution space may 
drastically change and even reduce to empty if the standard procedures of Calculus of Variations 
are not used.

\NewSection{The Relation between Higher Order Variations and Boundary Terms}

Let  us consider the following Lagrangian
$$
  L'(q,\dot q, \ddot q)=\dot q \ddot q + \Frac[1/2]\(\dot q^2-\om^2\>q^2\)+ \om^2 q\dot q
\fl{Lag1}$$
which is easily found to be equivalent to the Lagrangian of an harmonic oscillator  and to give 
rise to the same dynamics via Euler-Lagrange equations (of order 2).
Varying it we have
$$
\eqalign{
\de L'&= \de \dot q \ddot q + \dot q \de \ddot q+\dot q\>\de\dot q-\om^2\>q\>\de q +\om^2 \de q
\dot q+ \om^2 q\de \dot q=\cr
=& \frac[d/dt]\(\de q \ddot q\) - \de q \frac[d^3q/dt^3]
+\frac[d/dt]\(\dot q \de \dot q\) -\frac[d/dt]\(\ddot q \de q\)+ \frac[d^3q/dt^3]  \de q+\cr
&+\frac[d/dt]\(\dot q\>\de q\) -\ddot q\>\de q
-\om^2\>q\>\de q +\om^2 \de q \dot q
+\frac[d/dt]\(\om^2 q\de q\)-\om^2 \dot q\de q=\cr
=& \frac[d/dt]\( \dot q \de \dot q+\(\dot q +\om^2 q\)\de q\) -\(\ddot q +\om^2\>q\)\>\de q \cr
}
\fn$$
If following the standard prescriptions of Calculus of Variations we assume $\de q =0$ and $\de\dot
q=0$ on the boundary of an interval $[t_0, t_1]$ then we obtain in fact the equation of motion of
the $1d$-harmonic oscillator
$$
  \ddot q+\om^2\>q=0
\fl{1}$$%
This is no mistery since the Lagrangian \ShowLabel{Lag1} can be easily recasted as follows
$$
L'(q,\dot q, \ddot q)= \Frac[1/2]\(\dot q^2-\om^2\>q^2\)  +\Frac[d/dt]\(\Frac[1/2]\>\(\dot
q^2+\om^2 q^2\)\)
\fn$$
so that it manifestly differs from the harmonic oscillator Lagrangian $L(q,\dot q)=\Frac[1/2]\(\dot
q^2-\om^2\>q^2\)$ by a total time derivative (which is the mechanical equivalent of a pure
divergence term in field theory).  Hence, in this case, we know that the pure-divergence-term
$\Frac[d/dt]\(\Frac[1/2]\>\(\dot q^2+\om^2 q^2\)\)$ in the Lagrangian $L'$ is totally unessential
with respect to the equation of motion.  Let us stress that in this case the pure divengence term is
even zero on-shell because of the conservation of total energy, since the boundary term is nothing 
but the total derivative of the Hamiltonian.

If  one decides instead to fix only $\de q=0$ on the boundary, leaving $\de\dot q$ unfixed, then
extra boundary field equations are added 
in order to kill the extra boundary  contribution to the action. 
The equations of motion that follow form (2.2)
in this case are
$$
  \cases{ 
&  \ddot q+\om^2\>q=0\cr
 & \dot q_0=0}
\fn$$%
which in fact admit less solutions than Eq.  \ShowLabel{1}.  Notice that solutions to this problem
are in fact just a zero-measure set in the solution space of the $1d$-harmonic oscillator!

If one decides not to keep the first derivatives fixed, by adding pure divergencies one can even
invent nastier and nastier examples.  For instance, by considering the following 1-parameter family
of Lagrangians
$$
L''(q, \dot q, \ddot q; \La) =  \Frac[1/2]\(\dot q^2-\om^2\>q^2\)  +\Frac[d/dt]\(\Frac[1/6]\>\dot
q^3+\(\Frac[\om^2/2]\> q^2 +\La^2 \)\dot q\)
\fn$$
with $\La$ real,
which produce equations of motion in the form
$$
  \cases{ 
&  \ddot q+\om^2\>q=0\cr
 & \dot q^2_0 + \om^2 q^2_0=-\La^2}
\fn$$%
wee see that, for any $\La\not=0$, one has no solution at all, since there are no initial conditions
satisfying the boundary equation.  And even for $\La=0$ the solution space is much smaller than the 
solution space of the harmonic oscillator, since it reduces again to quiet.

\

\NewSection{Examples in GR}

Of course one could argue that field theory is not Mechanics and that in Field Theory there is more
space to play with.  Such an assumption is of course true, but still one has to pay a lot of
attention when playing\dots!  Let us then present similar situations in GR.

Let $M$ be a 4-dimensional manifold with boundary $\Om$ and let us consider the metric Lagrangian
$$
L= \sqrt{g} R -\na_\al \( \sqrt{g} g^{\mu\nu} \(u^\al_{\mu\nu}- \bar u^\al_{\mu\nu}\)\)
= \Big[\sqrt{g} g^{\al\be}(\Ga^\rho_{\al\si}\Ga^\si_{\rho\be}-\Ga^\si_{\si\rho}\Ga^\rho_{\al\be})
+d_\si (\sqrt{g} g^{\al\be}\bar u^\si_{\al\be})\Big]\> ds
\fl{l1o}$$
where: $ds$ is the standard local volume element induced by the coordinates; here and below,
$\Ga^\al_{\be\mu}$ are the coefficients of the Levi-Civita connection of the metric $g$; we set
$u^\la_{\mu\nu}= \Ga^\la_{\mu\nu}-\de^\la_{(\mu} \Ga^\al_{\nu)\al}$ and $\bar u^\la_{\mu\nu}= \bar
\Ga^\la_{\mu\nu}-\de^\la_{(\mu} \bar\Ga^\al_{\nu)\al}$ for any connection $\bar \Ga^\la_{\mu\nu}$
chosen at will on $M$.  $\Ga^\al_{\be\mu}$ as well as $u^\la_{\mu\nu}$ are functions of the first
derivatives of the field $g_{\mu\nu}$, while $\bar \Ga^\la_{\mu\nu}$ is just a ``fixed
parametrization'' i.~e.\ a non-dynamical background (as one could easily see by realizing that the
Euler-Lagrange equations of \ShowLabel{l1o} with respect to $\bar \Ga^\la_{\mu\nu}$ are identities).
As long as the background connection $\bar \Ga^\al_{\be\mu}$ is considered, one is free to
fix it at will: it can be a generic connection or the Levi-Civita connection of a background metric $\bar g$
(which could even have in principle a different signature) depending on the situation.

The Lagrangian \ShowLabel{l1o} is covariant and first order in $g_{\mu\nu}$; the connection
$\bar\Ga^\al_{\be\mu}$ is not subjected to any field equations so that it can be any connection both
{\it a priori\/} and {\it a posteriori\/} (we stress that connections exist globally on any
manifold); bulk field equations for $g$ are vacuum Einstein field equations.

The background $\bar u^\la_{\mu\nu}$ is here added to preserve covariance.  One could fix
coordinates so that $\bar u^\la_{\mu\nu}=0$ (usually at a point), or consider a fixed $\bar
u^\la_{\mu\nu}(x)$ as a point dependence (we stress that it is relegated into a divergence).  Our
procedure is analogous to the one used by Hawking and Ellis (see \ref{HawkingEllis}) to study the
Cauchy problem in Relativity; there a background (metric) is used at the level of field equations,
to show essential hyperbolicity,
while here it is used at the level of the action.  The two approaches are equivalent since the
background is non-dynamical and its fixing commutes with the derivation of field equations; see also
\ref{FFCov}.

The variation of this Lagrangian is given by
$$
\de L= \sqrt{g} G_{\mu\nu}\de g^{\mu\nu} -\na_\la \( \sqrt{g} (\de^\mu_{(\al}\de^\nu_{\be)} -
\frac[1/2] g_{\al\be}) (u^\la_{\mu\nu}-\bar u^\la_{\mu\nu})  \de g^{\al \be}- \sqrt{g}
g^{\mu\nu}\de \bar u^\la_{\mu\nu}\)
\fn$$
with $G_{\mu\nu}=R_{\mu\nu}-\frac[1/2] R g_{\mu\nu}$. Applying standard techniques of Calculus of Variation one 
obtains only the bulk standard field equations $G_{\mu\nu}=0$.
If, instead, one fixes
only $\de g^{\mu\nu}=0$ on the boundary, then a new boundary equation (associated to 
$\de \bar u^\la_{\mu\nu}$) is added
$$
\sqrt{g} g^{\mu\nu}|_\Om=0
\quad\then
g_{\mu\nu}|_\Om=0
\fn$$
This boundary condition is not only incompatible with the bulk field equations, but with kinematics in
the first place (metrics are assumed in fact to be non-degenerate so that they are everywhere
forbidden to vanish). 
Hence if one considers the Lagrangian \ShowLabel{l1o}, that differs from standard GR by a
divergence, and fixes the metric only, then the solution space is {\sl empty\/}!

One could argue that the background $\bar u^\la_{\mu\nu}$ is unphysical since it has no dynamics and
that therefore there is no need to consider its variations.  That is certainly reasonable though the
argument can be reversed: since the field $\bar u^\la_{\mu\nu}$ is unphysical, then physics should
be independent of how one decides to treat it: keeping it fixed or varying it, possibly varying an
underlying metric $\bar g_{\mu\nu}$ that fixes it on the boundary, alone or together with its first
derivative.  The above example shows instead how the physical predictions of the theory (in
particular the solution space) do depend on which unphysical degree of freedom is kept fixed on the
boundary.  Moreover, conservation laws would result to be affected by terms ensuing form the 
divergence (they can be easily calculated as in \ref{FFCov}).

Similar (but nastier) examples can be considered: e.g.\ the Lagrangian
$$
L'= \sqrt{g} R - \frac[1/\La]\na_\al \(\sqrt{g} g^{\mu\nu} R\(u^\al_{\mu\nu}- \bar
u^\al_{\mu\nu}\)\)
\fn$$
that is again classically equivalent to the Hilbert Lagrangian.
The variation is now
$$
\eqalign{
\de L'= &\sqrt{g} G_{\mu\nu}\de g^{\mu\nu} -\na_\la \(\frac[1/\La] \de( \sqrt{g} g^{\mu\nu})
R\(u^\la_{\mu\nu}- \bar u^\la_{\mu\nu}\) 
+ \frac[1/\La] \de R \sqrt{g} g^{\mu\nu} \(u^\la_{\mu\nu}- \bar u^\la_{\mu\nu}\) \)+\cr
& -\na_\la \(\frac[1/\La]\sqrt{g} g^{\mu\nu} (R -\La)\de u^\la_{\mu\nu}-  \frac[1/\La]\sqrt{g}
g^{\mu\nu} R\de \bar u^\la_{\mu\nu} \)
}
\fn$$

Here, if we fix $\de g^{\mu\nu}=0$  leaving $\de R$, $\de u^\la_{\mu\nu}$ and $\de \bar
u^\la_{\mu\nu}$ unconstrained on the boundary, we have three boundary field equations 
$$
\cases{
& \sqrt{g} g^{\mu\nu}\(u^\al_{\mu\nu}- \bar u^\al_{\mu\nu}\) \de R |_\Om=0 
\quad\then
u^\al_{\mu\nu} |_\Om= \bar u^\al_{\mu\nu}|_\Om\cr
&\sqrt{g} g^{\mu\nu} (R-\La)\de u^\la_{\mu\nu}|_\Om=0 
\quad\then R|_\Om=\La\cr
&g^{\mu\nu} R \de \bar u^\la_{\mu\nu} |_\Om=0 
\quad\then R|_\Om=0\cr
}
\fl{eq2}$$
As in the previous example, these three conditions are incompatible and the resulting solution space
is again empty.  Unlike the previous example, however, if in this case one decides not to vary the
background the first two equations in \ShowLabel{eq2} are still obtained along with Einstein
equation; they (in particular, the second one) are enough to force the solution space to be empty.
Here the troubles are generated exactly from not fixing $\de u^\la_{\mu\nu}$ at the boundary.  If
now one adds to the Lagrangian a divergence that suitably counterbalance the first constraint, then
this is enough, for any $\La\not= 0$, to prevent Minkowski spacetime from being a solution of the
theory, with a devastating effect on Newtonian limit and the physical interpretation of the whole
theory.)  The first condition imposes in fact to $g_{\mu\nu}$ an arbitrary asymptotic; if $\bar
u^\la_{\mu\nu}$ is suitably chosen, then one could impose to $g_{\mu\nu}$ to be asymptotically AdS,
dS or anything else.  In any case, the solution space is again {\it empty\/}!

Other even more complicated examples can be studied under the form
$$
L_f= \sqrt{g} R - \na_\al \(\sqrt{g} g^{\mu\nu} f(R; \La, \dots)\(u^\al_{\mu\nu}- \bar
u^\al_{\mu\nu}\)\)
\fn$$

We stress that of course there are reasonable boundary terms which do not force the solution space
to be empty, but there is no guiding principle helping one in distinguishing good boundary terms
from bad ones, so that such a procedure should be better avoided (being misleading) or, if really
necessary, treated with the correct mathematical instruments.  All this in the case that the
``real'' Lagrangian we start deforming is the Hilbert Lagrangian, that is known to be the only
non-trivial second order Lagrangian linear in the curvature of a metric field.  Linearity implies
Hamiltonian degeneracy, so that the second order theory is essentially equivalent to a first order
theory with second order field equations.  It is exactly this degeneracy and the existence of a
family of covariant first order  (see \ref{FFCov}) that allows one to play with a certain success with the addition of
divergencies.  One should be aware that such a method {\sl cannot\/} hold any longer in more general
families of gravitational theories, such as e.~g.\ all $f(R)$ , Gauss-Bonnet, Lovelock, Chern-Simons
Lagrangians and so on, including all effective Lagrangians that ensue from low limits of spacetime
and/or quantum requirements.

\ms

 \NewSection{Conclusions}

We have here considered two attitudes  in a variational principle of order $k$.
Let us summarize our point. 
A {\sl weakly critical configuration\/} is a configuration that extremizes the action for any
deformation which vanishes along the boundary (while the field derivatives are left unconstrained).

A {\sl critical configuration\/} is instead a configuration which extremizes the action for any deformation
which vanishes together with its derivatives (up to order $k-1$)
along the boundary.  

Of course a weakly critical configuration is also critical, while the converse is false in general.
From these simple examples we may easily conclude that, in a theory of order $k$,
pure-divergence-terms may be 
considered unessential with respect to the field equations {\sl only if\/} one considers critical
configurations.
On the contrary, by adding boundary terms to the action one can easily force the space of weakly
critical configurations to be smaller or even empty.

Of course one is free to abandon the invariance of the action with respect to boundary
contributions (as in a sense is done in the Hamiltonian formalism).
Unfortunately, such an attitude strongly impacts on conservation laws which are an essential part of the
physical interpretation of the theory as well.

Weakly critical configurations are considered in \ref{Wald} (against the standard results in
Variational Calculus and other important monographs in GR that more correctly consider only critical
configurations; see \ref{HawkingEllis}, \ref{Caroll}, \ref{tHooft}).  In our opinion there is no
real reason to impose an often artificial boundary term to a covariant action, breaking general
covariance, in order to allow more general deformations of fields.  Deformations in Lagrangian
formalism have indeed no physical meaning.  In Mechanics they are called in fact {\sl virtual\/}
dispacements also to stress the fact that they are not physical and they just need to be generically
independent.

Any procedure that fixes fileds and no derivatives at the boundary is certainly very similar (if not technically identical) to a gauge fixing. Gauge
fixing are useful in practice in special situations 
but there is no reason to break gauge covariance by fixing a gauge when a gauge covariant procedure
allows to obtain the same result from
a more fundamentally satisfactory point of view.

Another way of considering these examples is from control theory in the Hamiltonian framework.
Boundary terms of the action are exactly the way of mimiking control theory at the Lagrangian level.
In such a framework one is not concerned with computing physical configurations (namely, solutions
of field equations) but how (and whether) physical configurations can respond to some constraint
imposed at the boundary.  For example, computing the electric field in a space with a conductor,
knowing that the boundary, i.e.~the surface of the conductor, is equipotential.
 
In this context the extra boundary equations are exactly interpreted as the condition one wishes to
impose at the boundary.
Here (and only here) one should guarantee that the boundary conditions imposed can be physically
realized.
It is no surprise that in certain cases there exist no configuration obeying those boundary
conditions, meaning that one cannot physically impose those particular
boundary conditions.

We have to stress that in gravitational experiments we are now technologically unable to impose
{\sl any\/} boundary conditions. 
It is therefore interesting to know that some requirements are forbidden in principle.

We have also to stress that the framework of control theory is by no means related to the
determination of solutions of field equations, where by definition one wants to obtain {\sl all\/}
possible field configurations.  Moreover, if we {\sl unnecessarily\/} rely on boundary terms to
obtain field equations, then this freedom cannot be exploited to deal with conservation laws (see
\ref{Augmented}).  In fact, it is well--known that, although divergencies leave invariant critical
configurations, they affect conservation laws and conserved quantities that are an important part
of the physical interpretation of the model.  If boundary terms are fixed for field equations one
could only hope conservation laws to turn out to make sense.

\Acknowledgements

This work is partially supported by  the contribution of INFN (Iniziativa Specifica NA12) and the
local research founds of Dipartimento di Matematica of Torino University.

\ShowBiblio

\end